\documentclass[pra,a4paper,aps,twocolumn,superscriptaddress,floatfix]{revtex4-2}
\usepackage{natbib}
\usepackage{graphicx}% Include figure files
\usepackage{nicefrac}
\usepackage{dcolumn}% Align table columns on decimal point
\usepackage{bm}% bold math
\usepackage{multirow}
\usepackage[dvipsnames]{xcolor}
\usepackage{braket}
\usepackage{amsmath}
\usepackage{subcaption}
\usepackage{soul}

\begin{document}

\preprint{APS/123-QED}

\title{Experimental investigation of nonclassicality in the simplest scenario via the degrees of freedom of light}

\author{J. M. M. Gama} 

\affiliation{Programa de Pós-graduação em Física, Instituto de F\'isica, Universidade Federal Fluminense, Av. Gal. Milton Tavares de Souza s/n, Niter\'oi, RJ, 24210-340, Brazil}

\author{G. T. C. Cruz}

\affiliation{Programa de Pós-graduação em Física, Instituto de F\'isica, Universidade Federal Fluminense, Av. Gal. Milton Tavares de Souza s/n, Niter\'oi, RJ, 24210-340, Brazil}

\author{Massy Khoshbin}

\affiliation{Independent Researcher}

\author{Lorenzo Catani}

\affiliation{International Iberian Nanotechnology Laboratory (INL), Av. Mestre José Veiga s/n, 4715-330 Braga, Portugal}

\author{J. A. O. Huguenin}

\affiliation{Programa de Pós-graduação em Física, Instituto de F\'isica, Universidade Federal Fluminense, Av. Gal. Milton Tavares de Souza s/n, Niter\'oi, RJ, 24210-340, Brazil}

\affiliation{Instituto de Ciências Exatas, Universidade Federal Fluminense, R. Des. Ellis Hermydio Figueira, 783, Volta Redonda - RJ, 27213-415, Brasil}

\author{W. F. Balthazar}

\affiliation{Programa de Pós-graduação em Física, Instituto de F\'isica, Universidade Federal Fluminense, Av. Gal. Milton Tavares de Souza s/n, Niter\'oi, RJ, 24210-340, Brasil}

\affiliation{Instituto Federal do Rio de Janeiro,  R. Antônio Barreiros, 212, Volta Redonda - RJ, 27213-100, Brasil}

\date{\today}

%\affiliation{Instituto de Ci\^encias Exatas, Universidade Federal Fluminense, 27213-145, Volta Redonda, Rio de Janeiro, Brazil}
%\affiliation{International Iberian Nanotechnology Laboratory (INL), Av. Mestre José Veiga, 4715-330 Braga, Portugal.}
%\affiliation{Instituto Federal do Rio de Janeiro, 27213-100, Volta Redonda, Rio de Janeiro, Brazil}

\email{wagner.balthazar@ifrj.edu.br}

\begin{abstract}

In this work, we experimentally investigate the classical-light emulation of different notions of nonclassicality in the simplest scenario. We implement this prepare-and-measure scenario involving four preparations and two binary-outcome measurements using two distinct experimental setups that exploit different degrees of freedom of light: polarization and first-order Hermite-Gaussian transverse modes. We additionally model experimental noise through an all-optical setup that reproduces the operational effect of a depolarizing channel. Our experimental results are consistent with the findings of Khoshbin et al. [Phys. Rev. A 109, 032212 (2024)]: under the assumption that the two measurements performed form a tomographically complete set, the observed statistics violate their noise-robust inequalities, indicating inconsistencies with preparation noncontextuality and bounded ontological distinctness for preparations.
Although our implementation uses classical light, it reproduces the statistics predicted for the simplest scenario. Since the states and measurements of this scenario underpin computational advantages in tasks such as two-bit quantum random access codes---among the simplest communication primitives enabling semi-device-independent certification of nonclassicality---our implementation is directly relevant for such applications.% \ze{can be used in single-photon regime by essentially changing the light source and the detectors.  Therefore, has direct relevance for such applications.} \lorenzo{[I’d avoid adding this last sentence here, since it would require a proper discussion; otherwise it may trigger an immediate reaction along the lines of: “well, of course—if you change key components of the experiment (the light source and detectors), you can do many more things!"]}.

\end{abstract}

\maketitle

%\tableofcontents

\section{\label{sec:level1}Introduction}
Any satisfactory notion of nonclassicality must ultimately be subject to experimental scrutiny. In a recent work~\cite{Khoshbin2024}, noise thresholds have been derived for witnessing two key notions of nonclassicality in what is known as the \emph{simplest scenario}: preparation contextuality~\cite{Spekkens2005} and violations of BOD$_P$ (bounded ontological distinctness for preparations~\cite{Chaturvedi2020}). The simplest scenario consists of four preparations and two tomographically complete measurements \footnote{The assumption of tomographic completeness, which is crucial for discussing generalized contextuality, states that there are no operationally accessible measurements that can distinguish between preparations which are indistinguishable by the measurements included in the experimental set.}, as shown in Figure \ref{fig:noiseless_case}, and was first studied in~\cite{Pusey2018}. It is termed ``simplest'' because it represents the minimal nontrivial setting in which preparation contextuality can be demonstrated. 
Figure~\ref{fig:noiseless_case} illustrates the noiseless simplest scenario. In a realistic setting, however, the preparations do not exactly correspond to the pure states shown there, but are instead represented as in Figure~\ref{fig:noisy_case}.
In~\cite{Khoshbin2024}, three different approaches to testing nonclassicality in the latter scenario were examined: the method introduced by M.~Pusey~\cite{Pusey2018} and that developed by I.~Marvian~\cite{Marvian2020} for witnessing preparation contextuality, together with a novel approach for witnessing violations of BOD$_P$. A central result of that work is that these three approaches coincide in their ability to detect nonclassicality, provided that the level of experimental noise remains below a certain threshold, namely $\delta < 0.007$ (or $\delta < 0.02$ in the presence of quantum depolarizing noise). This result implies that experimenters have freedom in selecting the approach best suited to their experimental constraints, as long as the noise lies within this regime. Such flexibility is particularly important in scenarios where some approaches are impractical, for example in the noisy two-bit parity-oblivious multiplexing protocol~\cite{Spekkens2009}.

Preparation contextuality refers to the impossibility for a theory to admit of a preparation noncontextual ontological model~\cite{Spekkens2005}. This means that the theory predicts operational equivalences between certain preparations -- such as the two convex decompositions of pure states of a qubit in Fig.~\ref{fig:noiseless_case} that yield the maximally mixed state -- while nevertheless representing these preparations as distinct in the ontological model. The latter constitutes a formal way to provide a realist explanation of the predictions of a theory~\cite{Harrigan}. The assumption of noncontextuality itself is motivated by a methodological principle inspired by Leibniz’s principle of the identity of indiscernibles~\cite{SpekkensLeibniz}, or, equivalently, by the principle of no operational fine tuning~\cite{CataniLeifer2020}.  
Appealing to the same credentials, one may further expect that not only operational equivalences should be mapped to ontological identities, but also that \emph{operational differences} should be preserved at the ontological level. This requirement underlies the notion of bounded ontological distinctness~\cite{Chaturvedi2020}.

While violations of BOD$_P$ remain comparatively less explored, preparation contextuality has emerged as a leading notion of nonclassicality. First, it applies to a wide range of scenarios and, whenever these apply, it coincides with other notions of nonclassicality, such as the negativity of quasiprobability representations~\cite{Ferrie2008,Spekkens2008,SchmidKD2024} and Bell nonlocality~\cite{Bell1964,Wright2023}. Second, it allows one to identify what is precisely nonclassical in a variety of quantum phenomena, including state discrimination~\cite{Schmid2018,Jaehee2021,Flatt2022,Sumit2022,Roch2024,Flatt2025,Ying2025}, interference~\cite{ToyFieldTheory,Wagner2022,Catani2023WP}, uncertainty relations~\cite{Catani2022UR}, cloning~\cite{LostaglioSenno2020}, broadcasting~\cite{Jokinen2024}, compatibility~\cite{Tavakoli2020,Selby2023}, coherence~\cite{Wagner2022,Rossi2023}, metrology~\cite{lostaglio2020certifying}, thermodynamics~\cite{lostaglio2020certifying,Comar2025}, weak values~\cite{Pusey2014}, quantum Darwinism~\cite{Baldijao2021}, and communication tasks~\cite{Spekkens2009,Chailloux2016,SahaAnubhav2019,Yadavalli2022,Fonseca2025}. Moreover, preparation contextuality has been systematically studied within the frameworks of resource theories~\cite{Duarte2018,Gonda2024}, graph-theoretic approaches~\cite{Kunjwal2019,Kunjwal2020}, and has also been adopted in the analysis of Wigner’s-friend-type scenarios~\cite{Walleghem2025}. Third, preparation contextuality can be subjected to direct experimental tests~\cite{Mazurek2016,Zhan2017,Zhang2019,Mazurek2021,Giordani2023,FanQin2026}.

Regarding the empirical study of preparation contextuality, the first definitive test was reported in~\cite{Mazurek2016}, where a prepare-and-measure scenario involving six preparations and three binary-outcome measurements was implemented using the polarization of single photons. 
The resulting operational data were analyzed within the framework of generalized probabilistic theories (GPTs)~\cite{Janotta2014,Plavala2023}, in which three binary measurements are tomographically complete, leading to a clear violation of the noncontextual bound. 
Subsequent work by Zhan \textit{et al.}~\cite{Zhan2017} tested the generalized noncontextuality inequality derived in~\cite{Liang2011} using a similar photonic platform.
The resulting operational data demonstrated that the observed correlations cannot be accounted for by a generalized noncontextual model, even for a single-qubit system with unsharp measurements.
A related line of experiments, e.g., Zhang \textit{et al.}~\cite{Zhang2019}, explored contextuality tests in optical fields using different operational definitions of measurement events, comparing the manifestations of contextuality between single-photon systems and classical coherent states. Furthermore, later work by Mazurek \textit{et al.}~\cite{Mazurek2021} introduced techniques to mitigate the loophole associated with the assumption of tomographic completeness by constructing secondary preparations and measurements, and Giordani \textit{et al.}~\cite{Giordani2023} tested the application of contextuality witnesses to operational tasks identified in \cite{Wagner2022}.

Beyond tests based on genuinely quantum resources such as single-photon sources, a complementary line of work has explored how the same operational signatures of nonclassicality can be emulated and probed using classical optical fields. Among other applications, polarization and first-order transverse spatial modes are used to generate well-known maximally nonseparable spin-orbit modes \cite{topo,Balthazar2016}, employing classical light sources, which emulate the mathematical structure and correlation properties of maximally \cite{Pereira} and partially \cite{lamego2023partial} entangled states, with applications in Bell-like scenarios \cite{bell1, bell3}. In this sense, operational signatures of nonclassicality -- such as generalized contextuality -- can be investigated not only with single-photon sources but also using classical optical beams.

Spin-orbit modes are also employed as optical platforms to investigate features associated with environment-induced entanglement \cite{environ}, random walks \cite{lamego2024transition}, quantum cryptography \cite{ccrypt}, quantum games \cite{maioli2019quantization}, and, more recently, X-states and quantum discord \cite{Xstate, DiscordSpinOrbitModes}, as well as Kochen–Specker contextuality~\cite{li2017experimental} and Kujala-Dzhafarov contextuality-by-default~\cite{context}. More generally, polarization degrees of freedom have also been used to emulate Markovian~\cite{obando2020simulating} and non-Markovian~\cite{passos2019non} quantum decoherence, and simulate quantum thermal machines~\cite{quantThermoOPTSIM}.

Following this line of research, in this work we perform experiments in the simplest scenario shown in Fig.~\ref{fig:simplest_scenario}. We consider two degrees of freedom of classical light: polarization and first-order Hermite--Gaussian transverse modes. Under the assumption that the two implemented measurements form a tomographically complete set, we experimentally test the predictions of \cite{Khoshbin2024}. In particular, we verify the preparation contextuality witnesses introduced by Pusey and Marvian, and show that parity preservation, as defined in \cite{Khoshbin2024}, is violated, thereby implying violations of $\mathrm{BOD}_P$. To the best of our knowledge, this constitutes the first experimental investigation of a violation of $\mathrm{BOD}_P$.
We also study the case where noise is modeled via a depolarizing channel. Overall, our experimental results are in full agreement with the theoretical predictions. Here, tomographic completeness means that the two implemented binary-outcome measurements are sufficient to reconstruct the effective two-dimensional operational description of the preparations. In other words, within this reduced description, two suitably chosen binary measurements suffice to reconstruct the preparation statistics (as in rebit quantum theory). We discuss this assumption further in the conclusion, as well as how our experiment addresses other experimental idealizations~\cite{Mazurek2016}. 

The remainder of this article is organized as follows. Section~\ref{sec:level2} provides an overview of the results presented in \cite{Khoshbin2024} concerning preparation contextuality and violations of $\mathrm{BOD}_P$ in the presence of noise. Section~\ref{sec:experiment} details the optical experimental setup implementing the simplest scenario. Section~\ref{sec:results} presents the experimental results. Finally, Section~\ref{sec:conclusions} concludes with a discussion of the results and an outline of future research directions.

\section{\label{sec:level2} Theoretical Framework}

\begin{figure}[htbp]
    \centering

    \begin{subfigure}{\linewidth}
        \centering
        \includegraphics[width=\linewidth]{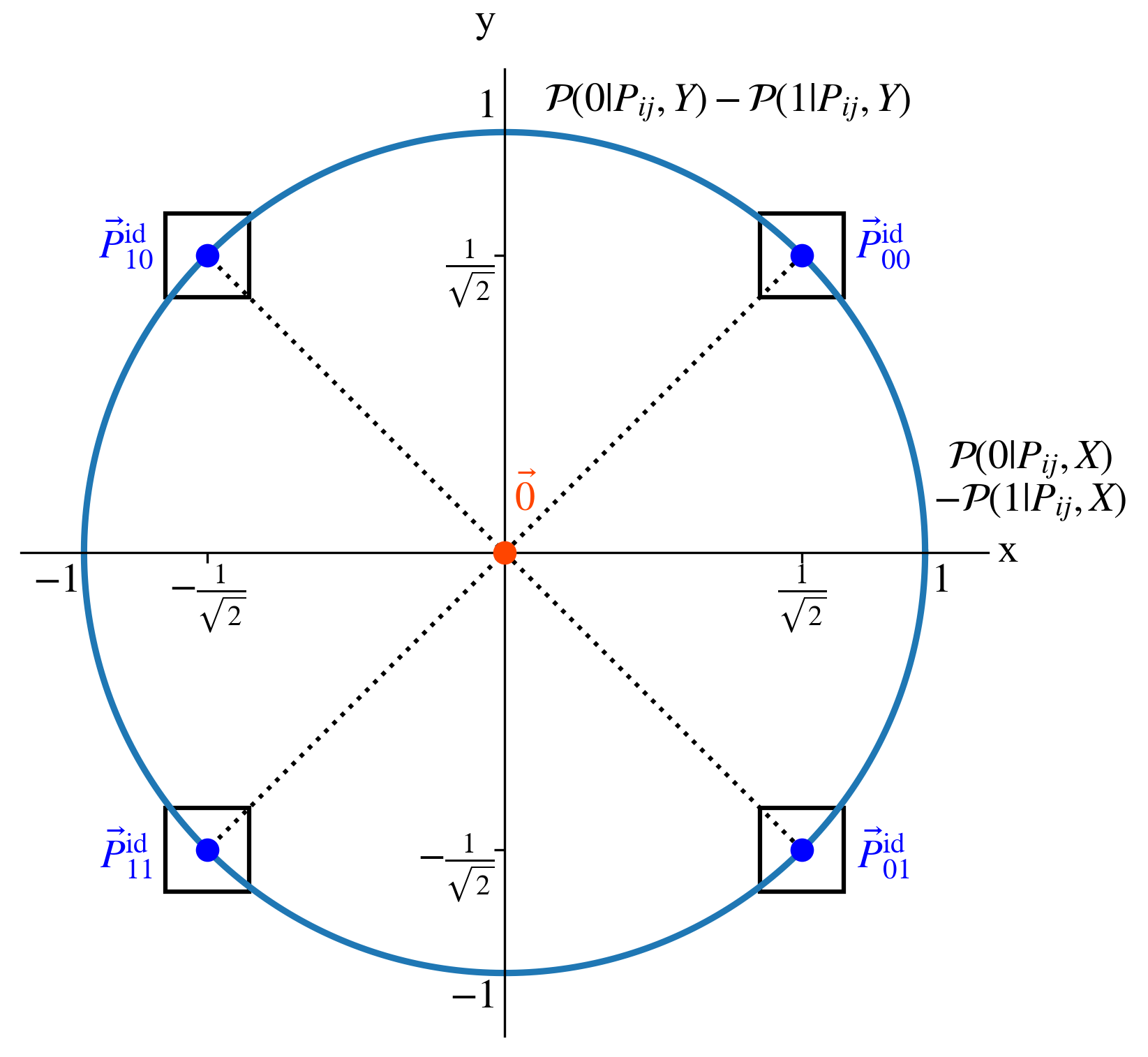}
        \caption{The simplest scenario in the noiseless case: four ideal preparations are shown on the boundary of the Bloch circle, along with two tomographically complete measurements represented by the $x$ and $y$ axes. The operational equivalence of Eq.\eqref{noiseless_equ} is indicated by the vector $\vec{0}$.}
        \label{fig:noiseless_case}
    \end{subfigure}

    \vspace{0.5cm}

    \begin{subfigure}{\linewidth}
        \centering
        \includegraphics[width=\linewidth]{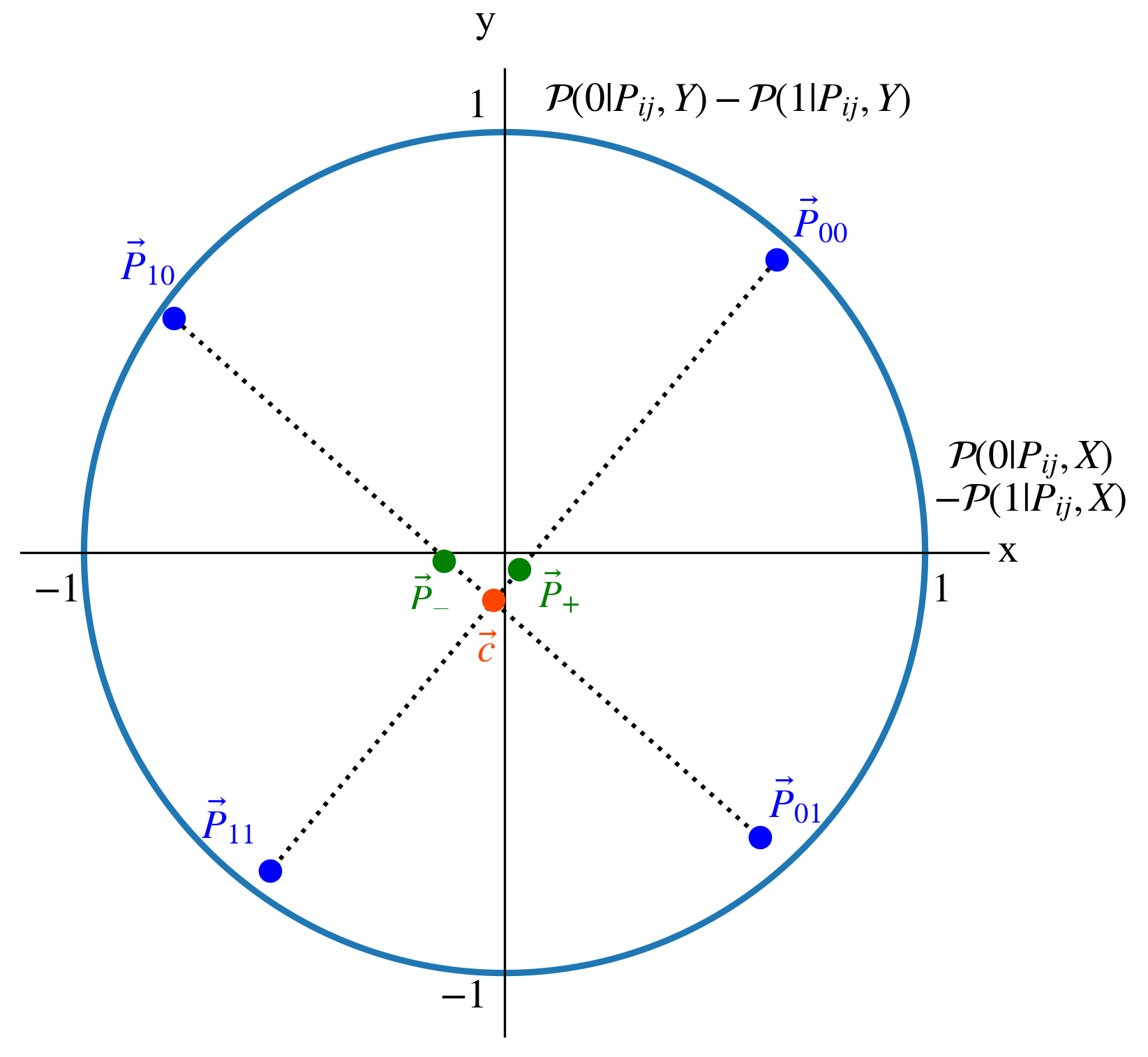}
        \caption{The simplest scenario in the noisy case: four noisy preparations are shown, satisfying the operational equivalence, $\vec{c}$, in Eq.~\eqref{op_equ}. The midpoints representing even and odd parity mixtures are indicated with $\vec{P_+}$ and $\vec{P_-}$, respectively.}
        \label{fig:noisy_case}
    \end{subfigure}

    \caption{Simplest scenario in the noiseless and noisy cases.}
    \label{fig:simplest_scenario}
\end{figure}

In this section, we summarize the results presented in \cite{Khoshbin2024}, in which a theoretical framework for testing two notions of nonclassicality, preparation contextuality and violations of bounded ontological distinctness for preparations, is provided. The focus is on the simplest non-trivial scenario, which consists of four preparations and two binary-outcome tomographically complete measurements. The preparations are denoted by $\{P_{ij} \}= \{ P_{00},  P_{01},  P_{10}, P_{11}\}$, and the tomographically complete measurements by ${X, Y}$. Each state $P_{ij}$ is represented by a vector $\vec{P}_{ij} = (x_{ij},y_{ij})$ in the $x$-$y$ plane of the Bloch sphere, as shown in Fig.~\ref{fig:simplest_scenario}. The vector coordinates are given by:
\begin{equation}\label{1}
	x_{ij} = \mathcal{P}(0|P_{ij}, X) - \mathcal{P}(1| P_{ij}, X),     
\end{equation}
\vspace{-0.8 cm}
\begin{equation}\label{2}
	y_{ij} = \mathcal{P}(0| P_{ij}, Y) - \mathcal{P}(1| P_{ij}, Y). 
\end{equation}

\noindent That is, the $x$-coordinate of $\vec{P}_{ij}$ is the difference between the probability of obtaining outcome $0$ and the probability of obtaining outcome $1$ when measuring $X$, and similarly with the $y$-coordinate for the measurement $Y$. It follows that two preparations $P_a$ and $P_b$ are operationally
equivalent, i.e., they yield the same outcome statistics for every measurement -- denoted by $P_a \simeq P_b$ -- if and only if the vectors representing their operational statistics are equal:
\begin{equation}
	P_a\simeq P_b \iff \vec{P}_a=\vec{P}_b.
\end{equation}

Within quantum theory, the following ideal preparations -- denoted $P^{\mathrm{id}}_{ij}$ -- yield the maximal violation of noncontextuality inequalities~\cite{Pusey2018} and, in particular, achieve the optimal probability of success in  2-bit parity-oblivious multiplexing~\cite{Spekkens2009}: 
\begin{equation}
	\begin{aligned}
		\vec{P}^{\text{id}}_{00} &= \left(\frac{1}{\sqrt{2}}, \frac{1}{\sqrt{2}}\right),   
		\vec{P}^{\text{id}}_{01} = \left(\frac{1}{\sqrt{2}}, - \frac{1}{\sqrt{2}}\right), \\
		\vec{P}^{\text{id}}_{10} &= \left(- \frac{1}{\sqrt{2}}, \frac{1}{\sqrt{2}}\right),  \vec{P}^{\text{id}}_{11} = \left(-\frac{1}{\sqrt{2}}, -\frac{1}{\sqrt{2}}\right).
	\end{aligned}
	\label{3}
\end{equation}

The above preparations form an operational equivalence in the completely mixed state, denoted by the vector $\vec{0}$ (see Fig. \ref{fig:noiseless_case}): 
\begin{equation}
\label{noiseless_equ}
\frac{1}{2} \vec{P}^{\mathrm{id}}_{00} + \frac{1}{2} \vec{P}^{\mathrm{id}}_{11} = \vec{0} = \frac{1}{2} \vec{P}^{\mathrm{id}}_{01}  + \frac{1}{2} \vec{P}^{\mathrm{id}}_{10}.
\end{equation}In more general realistic scenarios, the preparations of the simplest scenario form an operational equivalence in terms of the convex sum of the even and odd parity mixtures of the $P_{ij}$ (see Fig. \ref{fig:noisy_case}),
%It is important to discuss the notion of operational equivalence among the preparations. One relevant case is the convex sum of the even and odd parity mixtures of the $P_{ij}$, in which two operationally equivalent decompositions are given by:
\begin{equation}
	p \vec{P}_{00} + (1-p) \vec{P}_{11} = \vec{c} = q \vec{P}_{01}  + (1-q) \vec{P}_{10}, 
	\label{op_equ}
\end{equation}

\noindent where $p, q \in [0,1]$. In what follows, we will refer to the even and odd parity mixtures with weights $p=q=\frac{1}{2}$ and denote them as $P_+ = \frac{P_{00}+P_{11}}{2}$ and $P_- = \frac{P_{01}+P_{10}}{2}$, respectively.

In \cite{Khoshbin2024}, the authors characterize three distinct approaches to testing nonclassicality in the simplest scenario. Two follow the works developed by Pusey \cite{Pusey2018} and Marvian \cite{Marvian2020} on preparation contextuality, while the third is a novel approach for witnessing violations of BOD$_P$ \cite{Chaturvedi2020}. Each approach yields an inequality parameterized by a noise parameter $\delta$, allowing one to determine a noise threshold below which the inequality is violated and nonclassicality is thereby witnessed. The noise parameter $\delta$ represents the operational distance between the experimental (noisy) preparations, $P_{ij}$, and their ideal counterparts, $P_{ij}^{\text{id}}$:
\begin{equation}
	d(P_{ij},P^{\text{id}}_{ij}) = \frac{1}{2}\max{\{\left| x_{ij} - x^{\text{id}}_{ij} \right|,\left| y_{ij} - y^{\text{id}}_{ij} \right|\}}.
	\label{distance}
\end{equation}
We now present the inequalities in each approach, and their corresponding thresholds of violation.

\subsection{Pusey's approach}

In exploiting a connection with the CHSH scenario, in \cite{Pusey2018} Pusey derives a robust noncontextuality inequality which, in terms of $(x_{ij}, y_{ij})$, reads as:
\begin{equation}
    \begin{split}
        S(x_{ij},y_{ij}) = p(x_{00}+y_{00}+x_{11}+y_{11}) \\ 
        +q(x_{01}-y_{01}+x_{10}-y_{10})+  \\
        (y_{10}-x_{10}-x_{11}-y_{11})-2 \leq 0.
    \end{split}
    \label{Pusey2}
\end{equation}

In the case of quantum theory, this inequality has a maximum violation of $\approx 0.82$ using the preparations described in Eq.~\eqref{3}.

In \cite{Khoshbin2024}, the authors bound Eq.~\eqref{Pusey2} as a function of the noise parameter $\delta$ as follows:
\begin{equation}
    \begin{split}
        S(x_{ij},y_{ij})  \geq 2\sqrt{2} - 2 - 16\delta + 32 \sqrt{2}\delta^2.
    \end{split}
    \label{Pusey3}
\end{equation}

As a consequence, a noise threshold is obtained for Pusey's preparation noncontextuality inequality to be violated: if $d(P_{ij},P^{\text{id}}_{ij}) \leq 0.06$, then $S(x_{ij},y_{ij}) > 0$. 

\subsection{Marvian's approach}

Marvian's work quantifies preparation contextuality in terms of ``inaccessible information", which measures the amount of information that is not preserved from the ontological to the operational level \cite{Marvian2020}. To this end, a quantity is defined as the largest distance between pairs of distributions, $\mu_a, \mu_b$, corresponding to two operationally equivalent preparations $P_a$, $P_b$, minimized over all possible ontological models, as follows:
\begin{equation}\label{8}
C_{\text{prep}}^{\text{min}} \equiv \underset{\text{Models}}{\inf}  \underset{P_a \simeq P_b}{\sup} \, d(\mu_a, \mu_b),
\end{equation}

\noindent where the ontological distance $d(\mu_a, \mu_b)$ is quantified by the total variational distance between probability distributions: $\frac{1}{2}
\sum
\left| \mu_a - \mu_b \right|$. If $C_{\text{prep}}^{\text{min}} = 0$, then there exists an ontological model such that $d(\mu_a, \mu_b)=0$ anytime $P_a \simeq P_b$, which is the definition of preparation noncontextuality. However, if $C_{\text{prep}}^{\text{min}} > 0$, any ontological model must assign distinct distributions over the ontic states to certain pairs of operationally equivalent preparations, thereby exhibiting preparation contextuality.

In the simplest scenario, the authors in \cite{Khoshbin2024} bound this quantity in terms of the noise parameter $\delta$ as follows:
\begin{equation}\label{marvian}
C_{\text{prep}}^{\text{min}} \geq \frac{\sqrt{2}-4\delta - 1}{4(\sqrt{2}-4\delta)}.
\end{equation}

Consequently, one obtains a noise threshold for Marvian's preparation noncontextuality equality to be violated: if $d(P_{ij},P^{\text{id}}_{ij}) \leq 0.1$, then $C_{\text{prep}}^{\text{min}} > 0$.

\subsection{Approach witnessing violations of BOD$_P$}

The notion of bounded ontological distinctness for preparations demands that the difference between the operational distinguishability $s^{P_a,P_b}_{\mathcal{O}}$ of any two preparations $P_a,P_b$, and the corresponding ontological distinctness  $s^{\mu_a,\mu_b}_{\Lambda}$ is zero \cite{Chaturvedi2020}. In \cite{Khoshbin2024}, these two quantities are defined, in the case of the simplest scenario, in terms of the operational and ontological distances:
\begin{equation}
%s_{P_a,P_b} \equiv \frac{1}{2} \max_{M} \left\{ P(0 \mid P_a, M) + P(1 \mid P_b, M) \right\}. 
s^{P_a,P_b}_{\mathcal{O}} = \frac{1 + d(P_a,P_b)}{2}, \hspace{.2cm}
%s_{\mu_a,\mu_b} \equiv \frac{1}{2}\sum_{\lambda}\max\\mu_a(\lambda),\, \mu_b(\lambda)\}
s^{\mu_a,\mu_b}_{\Lambda} = \frac{1 + d(\mu_a,\mu_b)}{2}.
\label{bod_p}
\end{equation}

In this case, BOD$_P$ implies $d(\mu_a,\mu_b)-d(P_a,P_b)=0$. Therefore, while BOD$_P$ implies preparation noncontextuality, the converse does not hold in general.

If an ontological model does not satisfy BOD$_P$, then $d(\mu_a,\mu_b)-d(P_a,P_b) > 0$. To characterize violations of BOD$_P$, the quantity $\mathcal{D}_{P_a,P_b} \equiv d(\mu_a,\mu_b) -  d(P_a,P_b)$ is defined, with $\mathcal{D}^{\min}_{P_a,P_b}$ being its minimization over all possible ontological models. Using Eq.~\eqref{bod_p}, we obtain 
\begin{equation}
s^{\mu_a,\mu_b}_{\Lambda} - s^{P_a,P_b}_{O} = \frac{1}{2}\mathcal{D}_{P_a,P_b}.
\end{equation}

Thus, the requirement of BOD$_P$ for an operational theory is equivalently expressed as $\mathcal{D}^{\min}_{P_a,P_b} = 0$ $ \;\forall P_a,P_b$. 

%\subsection{Parity preservation }

As argued for in \cite{Khoshbin2024}, the special instance of BOD$_P$ for even and odd parity mixtures, $P_+ = \frac{P_{00}+P_{11}}{2}$ and $P_- = \frac{P_{01}+P_{10}}{2}$, is of particular interest. If $\mathcal{D}_{P_+,P_-} = 0$ holds for an ontological model, then we say that it satisfies \textit{parity preservation}. If $\mathcal{D}_{P_+,P_-}^{\min} > 0$ holds for an operational theory, then it does not admit of a parity preserving ontological model. Given that parity preservation is a special case of BOD$_P$, a violation of parity preservation is automatically a violation of BOD$_P$.   

In \cite{Khoshbin2024}, the authors relate a violation of parity preservation with a violation of preparation noncontextuality through the following implication:
\begin{equation}
C_{\mathrm{prep}}^{\mathrm{min}} 
>
\frac{2\left(1 + 2\sqrt{3}\right)\delta - 4\sqrt{2}\,\delta^{2}}
{1 - 2\sqrt{2}\,\delta}
\;\Rightarrow\;
\mathcal{D}_{P_{+},P_{-}}^{\min} > 0.
\label{parity:a}
\end{equation}

Combining the above result with that of Eq.~\eqref{marvian} gives rise to a threshold of $\delta \leq 0.007$ for violating parity preservation. Comparing the results of this section, we conclude that all three approaches to classicality are violated if the noise bound satisfies $\delta \leq 0.007$.

\subsection{Quantum depolarizing noise} \label{section_theo_depol}

A common way to model experimental noise is through the depolarizing channel, whereby each noisy preparation is described as a convex mixture of the ideal preparation $P_{ij}^{\mathrm{id}}$ and the completely mixed state $\frac{I}{2}$. In this case, an improved lower bound to Pusey's expression can be found so that Eq.~\eqref{Pusey3} is updated to the following \cite{Khoshbin2024}:
\begin{equation}
S(x_{ij}, y_{ij}) \ge 2\sqrt{2} - 2 - 8\delta + \frac{ 8\sqrt{2}\,\delta^{2} - 4\delta
}{
1 - \sqrt{2}\,\delta
}.
\label{pusey_depol}
\end{equation}

It follows that if $d(P_{ij},P^{\text{id}}_{ij}) \leq 0.07$, then $S(x_{ij},y_{ij}) > 0$. We therefore have an improved threshold of violation for Pusey's inequality when noise is modeled via the depolarization channel. 

Similarly, under depolarizing noise, Eq.~\eqref{parity:a} takes the following form:
\begin{equation}
C^{\min}_{\mathrm{prep}} > \delta +
\frac{ \sqrt{2}\,\delta}{1 - 2\sqrt{2}\,\delta}
\;\Rightarrow\;
\mathcal{D}^{\min}_{P_{+},P_{-}} > 0.
\label{marvian_depol}
\end{equation}

\begin{figure}[h!]
    \centering
	\includegraphics[width=1\linewidth]{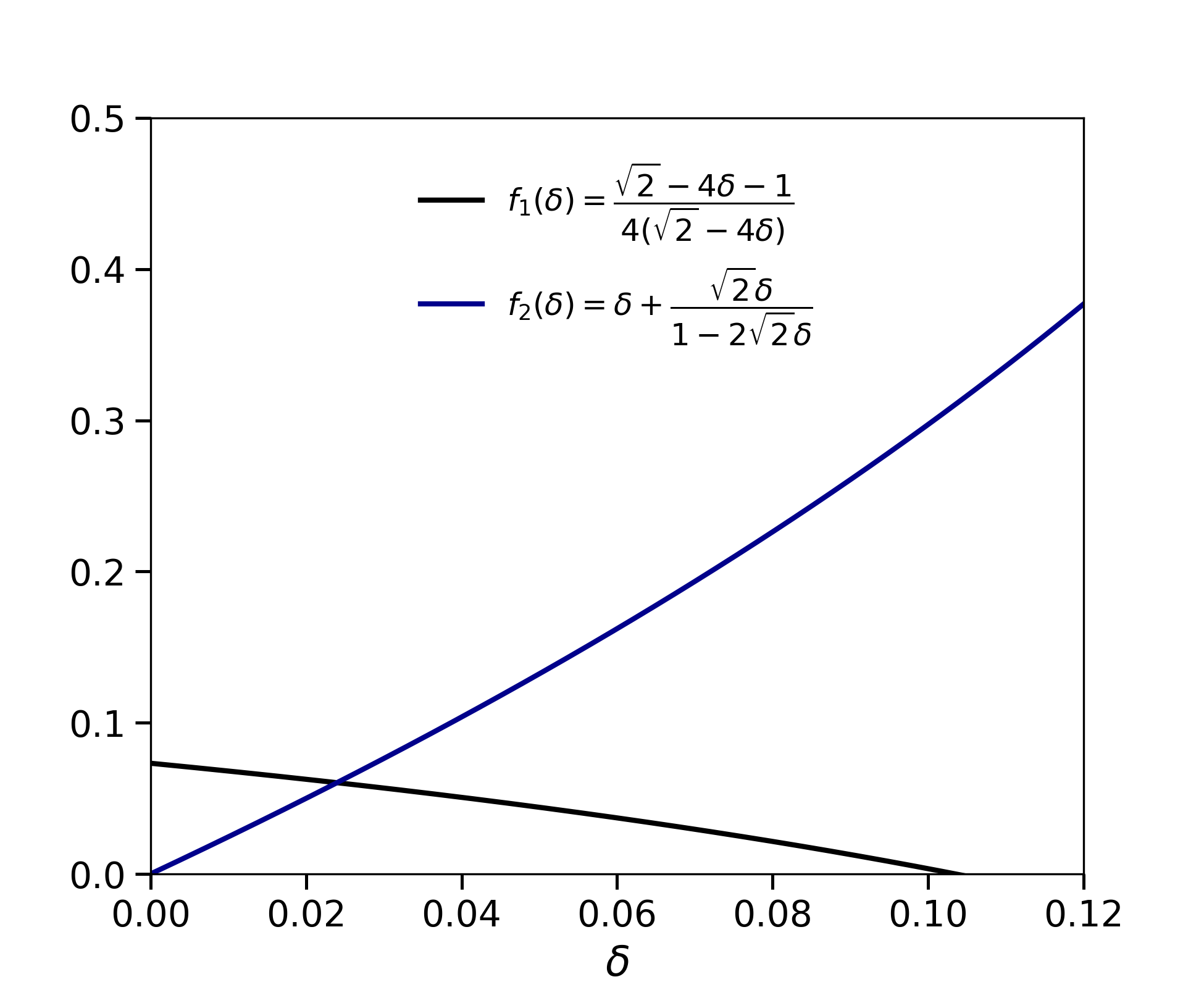}
	%\legend{Texto da legenda. (opcional)}
    \caption{Violation of parity preservation. The black curve indicates the bound in Eq.~\eqref{marvian} while the blue curve indicates the bound in Eq.~\eqref{marvian_depol}. A sufficient condition for $\mathcal{D}^{\min}_{P_{+},P_{-}} > 0$ is given by $f_1(\delta) > f_2(\delta)$, which holds for $\delta\leq 0.02$.} 
    \label{new_marvian}
\end{figure}

Combining this result with that of Eq.~\eqref{marvian} yields an improved noise threshold for violations of parity preservation, namely $\delta \leq 0.02$. Fig.~\ref{new_marvian} illustrates this condition.

We conclude that, under quantum depolarizing noise, all three approaches to classicality are violated whenever the noise bound satisfies $\delta \leq 0.02$.

\section{Experimental setups}
\label{sec:experiment}
In this section, we present the theoretical description of the experiments, which are implemented using an all-optical setup.
To prepare the four states described in Eq.~\eqref{3} and the two tomographically complete measurements, we employ polarization and first-order Hermite-Gaussian degrees of freedom of light with an intense laser beam. In addition, we use a linear circuit to emulate a depolarizing channel that models noise following the same setup as in \cite{Tiago2025}. In the experimental description below, we use standard quantum-mechanical terminology, as is customary in this type of experiment. However, the data analysis does not assume the validity of quantum theory, relying only on operational statistics and the assumption of tomographic completeness, in line with Ref.~\cite{Khoshbin2024}.  
 All experiments were performed using a diode-pumped solid-state laser at $532\,\mathrm{nm}$, with a power of $10\,\mathrm{mW}$ and a horizontally-polarized beam. The intensity was controlled using neutral density filters to avoid saturation of the detection system, and charge-coupled device (CCD) cameras were used for detection.

\subsection{Experiment with polarization}\label{exp_pol}

\begin{figure}[h!]
    \centering
	\includegraphics[width=1\linewidth]{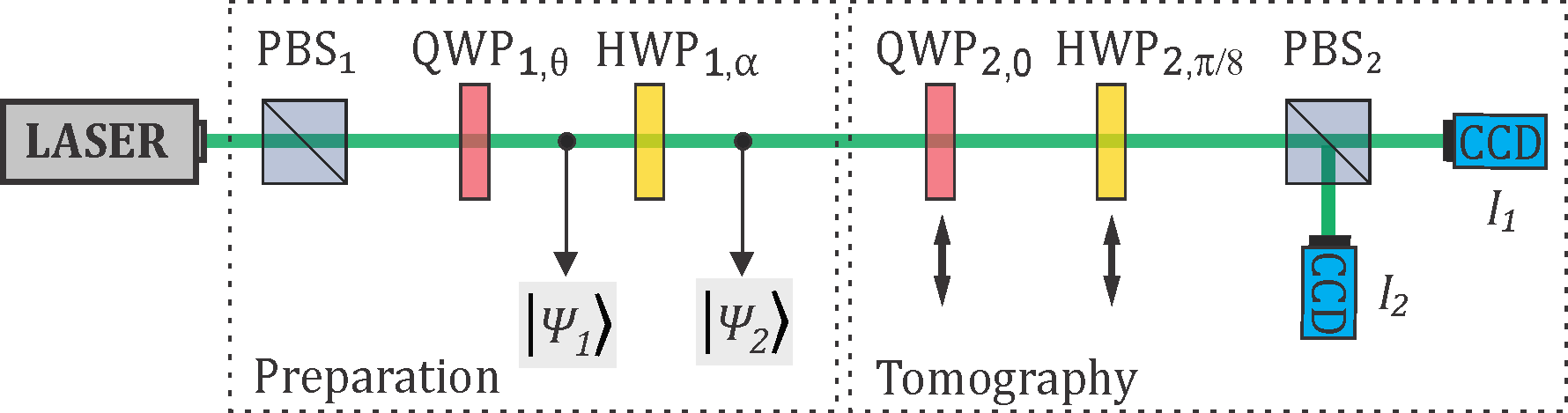}
	%\legend{Texto da legenda. (opcional)}
    \caption{Experimental setup for state preparation and measurement using polarization. %\lorenzo{[ Should we write few words highlighting why the angles for the preparations vary, while the angles for measurements are fixed?
    } 
    \label{Exp_1}
\end{figure}

For the polarization degree of freedom of light, the two-level system is encoded in the horizontal-vertical $\{H,V\}$ basis, %with
where we use the quantum bra-ket notation to denote its corresponding states,
$\ket{H} \equiv \ket{0}$ and $\ket{V} \equiv \ket{1}$. Fig.~\ref{Exp_1} illustrates the experimental optical setup divided into two blocks: preparation and tomography. 

Let us start by considering the preparation $\vec{P}^{\text{id}}_{00} = \left(\frac{1}{\sqrt{2}}, \frac{1}{\sqrt{2}}\right)$ in Eq.~\eqref{3}. We denote the corresponding state 
in bra–ket notation 
by $\ket{\psi_{00}}$. Initially, the horizontal polarization state $\ket{\psi} = \ket{H}$ is transmitted by the polarized beam splitter, $\textrm{\textrm{PBS}}_1$, and proceeds in the direction of the first quarter-wave plate oriented at an angle of $\theta=\pi/8$ relative to the horizontal axis, $\textrm{QWP}_{1,\nicefrac{\pi}{8}}$.  The state after $\textrm{\textrm{QWP}}_{1,\pi/8}$ is  $\ket{\psi_1} = U_{\textrm{QWP}_{1,\pi/8}}\ket{H}$, and, using Jones matrices \cite{jones1941new}, the state after the quarter-wave plate can be written as  $\ket{\psi_1} =  (0.8536 + 0.1466i) \ket{H} + (0.3535-0.3535i)\ket{V}$. 
This state passes through a half wave plate rotated at an angle of $\alpha = 3\pi/16$ relative to the horizontal axis, $\textrm{HWP}_{1,3\pi/16}$, leading to the state $\ket{\psi_2} = (0.6533-0.2706j)\ket{H} + (0.6533+0.2706j) \ket{V}$, which equals $\ket{\psi_{00}}$ up to transformations that do not affect measurement statistics, $\ket{\psi_{2}} = e^{0.3927i}\ket{\psi_{00}}$.

The second block in Fig.~\ref{Exp_1} is the complete polarization tomography \cite{altepeter2005photonic}, which implements what we denoted as the $X$ and $Y$ measurements in Section~\ref{sec:level2} and allows us to obtain the actual coordinates of the desired state vectors. The $\textrm{PBS}_2$ performs a measurement in the $\{H, V\}$ basis. The $\textrm{HWP}_2$ rotated at an angle of $\pi/8$ together with $\textrm{PBS}_2$ measure the diagonal and anti-diagonal polarization components -- the $\{D, A\}$ basis -- which corresponds to the $X$ measurement. The $\textrm{QWP}$ at an angle of $0$ together with $\textrm{HWP}_2$ and $\textrm{PBS}_2$ implement a measurement in the circular $\{R, L\}$ basis, which corresponds to the $Y$ measurement.

The intensities of the laser beam after each measurement are captured by a CCD camera and recorded in a single image. The normalized intensities $I_{m}/I_t$, with $m = 1,2$, $I_t = I_1 + I_2$, and where $m$ denotes the possible output ports $1$ and $2$, are interpreted as the probabilities $\mathcal{P}_1$ and $\mathcal{P}_2$ associated with the Stokes parameters. Therefore, when the state $\ket{\psi_{00}}$ is projected on the $\{H, V\}$ basis, the probabilities are $\mathcal{P}_H = 0.500$ and $\mathcal{P}_V = 0.500$, and the Stokes parameters are $S_0 = \mathcal{P}_H + \mathcal{P}_V = 1.000$ and $S_3 = \mathcal{P}_D - \mathcal{P}_A = 0.000$. The projections of $\ket{\psi_{00}} $ on the $\{D, A\}$ and $\{R, L\}$ basis are $\mathcal{P}_D = 0.854$, $\mathcal{P}_A = 0.156$, $\mathcal{P}_R = 0.854$ and $\mathcal{P}_L = 0.146$. Consequently, $S_1 = \mathcal{P}_D - \mathcal{P}_A = 0.707$ and $S_2 = \mathcal{P}_R - \mathcal{P}_L = 0.707$, which are the coordinates of the vector $\vec{P}^{\text{id}}_{00}$. 

\subsection{Experiment with transverse modes}\label{exp_TM}

\begin{figure}[h!]
    \centering
    \includegraphics[width=.97\linewidth]{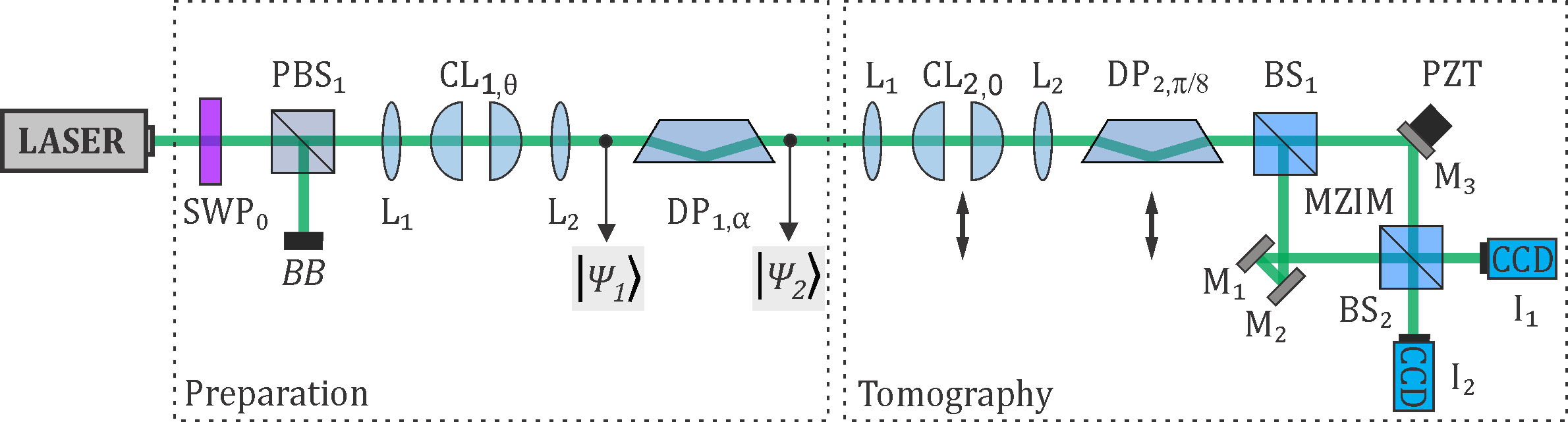}
	%\legend{Texto da legenda. (opcional)}
    \caption{Experimental setup for state preparation and measurement using transverse modes.} 
	\label{Exp_2}
\end{figure}

We now perform the experiment using first-order Hermite-Gaussian transverse modes with horizontal ($\ket{HG_{01}}$) and vertical ($\ket{HG_{10}}$) spatial orientation, which we associate with the states $\ket{h} \equiv \ket{0}$ and $\ket{v} \equiv \ket{1}$, respectively.
Fig.~\ref{Exp_2} shows the experimental setup to prepare the states defined in Eq.~\eqref{3} and to implement the $X$ and $Y$ measurements with these modes. 
The beam then passes through a \textrm{PBS}, which transmits the component $\ket{Hh}$ (horizontal polarization and horizontal mode), used as the initial state, and reflects the component $\ket{Vv}$ (vertical polarization and vertical mode), which is blocked. This method provides a simple and efficient way to prepare the spatial state $\ket{h}$ with fixed horizontal polarization, which remains unchanged throughout the experiment.
In analogy with the experiment exploiting the polarization degree of freedom, the cylindrical lens $\textrm{CL}_{1,\theta}$ and the Dove prism $\textrm{DP}_{1,\alpha}$ play the roles of $\textrm{QWP}_1$ and $\textrm{HWP}_1$, respectively. The lenses $L_1$ and $L_2$ form a telescope that is essential for mode matching.
By using the same angles as in the polarization experiment, we obtain the state
$\ket{\psi_2} = e^{0.3927 i}\ket{\psi_{00}}$,
which is physically equivalent to $\ket{\psi_{00}}$ up to transformations that do not affect measurement statistics. To prepare the remaining states in Eq.~\eqref{3}, namely $\ket{\psi_{01}}, \ket{\psi_{10}}$, and $\ket{\psi_{11}}$, we rotate $\textrm{CL}_1$ and $\textrm{DP}_1$ by setting the angles $(\theta,\alpha)$ to $(\pi/8,-\pi/16)$, $(-\pi/8,\pi/16)$, and $(-\pi/8,5\pi/16)$, respectively. We note that the Jones matrices of $\textrm{CL}_1$ and $\textrm{DP}_1$ coincide with those of $\textrm{QWP}$ and $\textrm{HWP}$, respectively.

The tomography block in Fig.~\ref{Exp_2} operates on the same principle as in the polarization measurement. 
The Mach-Zehnder interferometer with one additional mirror (\textrm{MZIM}) enables measurements in the $\{h,v\}$ basis, 
exploiting the fact that the polarization degree of freedom is uniform across all Hermite-Gaussian modes. 
A piezoelectric ceramic (\textrm{PZT}) is used to stabilize the phase difference between the interferometer arms at zero.
The combination of $\textrm{DP}_2$, rotated by $\pi/8$, and the $\textrm{MZIM}$ performs a measurement 
in the diagonal and anti-diagonal polarization basis $\{D,A\}$, which corresponds to the $X$ measurement. Similarly, $\textrm{CL}_{1,0}$ in conjunction with 
$\textrm{DP}_2$ and the \textrm{MZIM}, implements a measurement in the circular basis $\{R,L\}$, which corresponds to the $Y$ measurement. 
In this case with first-order
Hermite-Gaussian beam degree of freedom of light, the intensities at the output ports $I_1$ and $I_2$ correspond to the probabilities of the projective measurements 
$(p_h, p_D, p_R)$ and $(p_v, p_A, p_L)$, respectively. The theoretical predictions for these measurements 
and the corresponding Stokes parameters are identical to those obtained in the experiment with polarization degree of freedom of light.

\subsection{Experiment under depolarizing noise}\label{exp_dep_channel}

\begin{figure}[h!]
    \centering
	\includegraphics[width=1\linewidth]{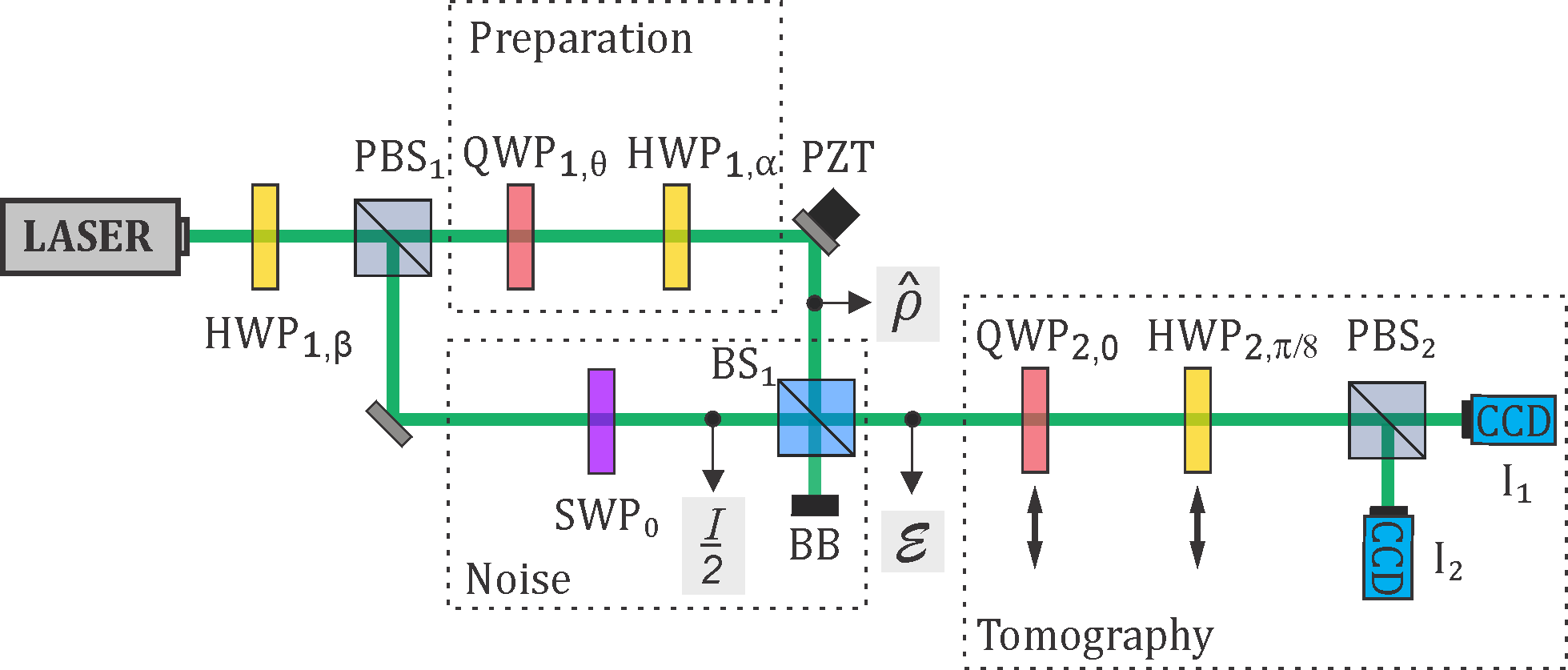}
	%\legend{Texto da legenda. (opcional)}
    \caption{Experimental setup to model noise via a depolarizing channel.} 
	\label{Exp_3}
\end{figure}

In this subsection, we assume that the experimental noise is modeled by a quantum depolarizing channel. %This means that the prepared states in Eq.~\eqref{3}, $\vec{P}_{ij}$, can be expressed as a convex combination of the corresponding ideal states $P^{id}_{ij}$ and the maximally mixed state.
Fig.~\ref{Exp_3} shows the optical setup implementing the depolarizing channel, based on \cite{Tiago2025}.

A laser beam prepared in the horizontally-polarized state $\ket{\psi} = \ket{H}$ passes through a half-wave plate rotated by an angle $\beta$, denoted \(\textrm{HWP}_{1,\beta}\). Initially, $\beta = 0$, so that all light is transmitted through $\textrm{PBS}_1$. We use the same optical circuit as shown in Fig.~\ref{Exp_1} for the state preparation. After $\textrm{BS}_1$, the state is described by the density operator $\rho = \ket{\psi}\bra{\psi}$ and the beam in the state $\mathcal{E}=\rho$ subsequently enters the tomography block, which has already been described in the previous case.
When $\beta \neq 0$, $\textrm{HWP}_{1,\beta}$ enables control of the relative intensities in the two polarization–path components. In this case, the vertically polarized component is reflected by $\textrm{PBS}_1$ and subsequently impinges on the spiral wave plate $\textrm{SWP}_0$, with its optical axis aligned with the horizontal polarization.
The vertically polarized component passing through \(\textrm{SWP}_0\) prepares the optical field in a radially polarized state, corresponding to a first-order polarization vortex,
$\ket{\Psi_+} = \frac{1}{\sqrt{2}}\left( \ket{Hh} + \ket{Vv} \right)$,
which is formally analogous to a Bell state in the tensor product of polarization and spatial-mode degrees of freedom.
In the single-qubit noise experiments considered here, genuine non-separability between these degrees of freedom is not required. Instead, the polarization–spatial structure of $\ket{\Psi_+}$ is exploited to implement the maximally mixed state $\tfrac{I}{2}$ for the spatial subsystem. This arises because the polarization degree of freedom is effectively traced out due to interference at $\textrm{BS}_1$ followed by projective measurements in the polarization basis.
In the upper arm, the beam remains in the fundamental Gaussian mode $HG_{00}$, whereas in the lower arm it occupies the first-order Hermite--Gaussian modes $HG_{01}$ and $HG_{10}$. Consequently, after the beam splitter $BS_1$, the resulting state is
\begin{equation}
\mathcal{E}(a) = (1-a) \rho +  a \frac{I}{2},
\label{eq::depolarizing}
\end{equation}

\noindent where $\rho = \ket{\psi}\bra{\psi}$ and $a \in [0,1]$ is related to the intensity -- the relative population of horizontally and vertically polarized photons in the beam. In the simplest case, increasing $a$ corresponds to adding noise to the prepared state. Geometrically, this is reflected in a shrinkage of the Bloch vector towards the center of the Bloch disk.

\section{Results}
\label{sec:results}

In this section, we analyze the experimental data that allow us to test the different approaches to witnessing nonclassicality described in Section~\ref{sec:level2}. %prepared contextual states by comparing different approaches to noncontextual inequalities. 

\subsection{Experiment with polarization}

\begin{figure}[!ht]
    \centering
    \includegraphics[width=0.67\linewidth]{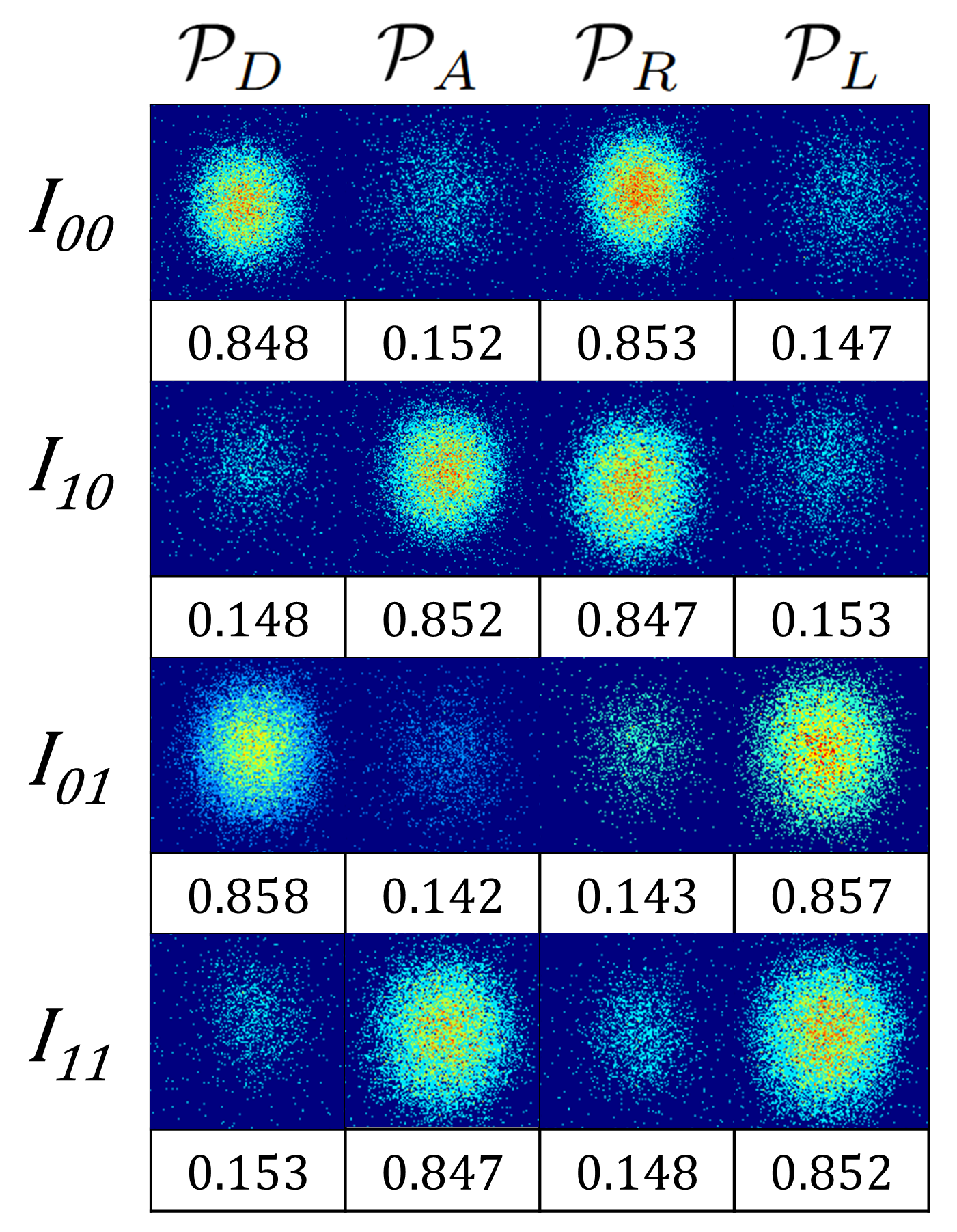}
    \caption{Output images (false color) for prepared polarization states.}
    \label{fig:output_1}
\end{figure}

Using the optical setup described in Section~\ref{exp_pol}, we prepare all states $P_{ij}$ of the simplest scenario, aiming to approximate the ideal states $P^{\mathrm{id}}_{ij}$ of Eq.~\eqref{3} as closely as possible. Fig.~\ref{fig:output_1} shows the experimental output intensities for the experiment with polarization. 
The rows label the intensities $I_{ij}$ for each prepared state, while the columns label the probabilities associated with the measurements on the diagonal and circular bases. The probabilities $\mathcal{P}_D$ and $\mathcal{P}_A$ ($\mathcal{P}_R$ and $\mathcal{P}_L$) are obtained from the intensities measured at output ports, $I_1$ and $I_2$, respectively.
For example, $I_{00}$ is associated with the 
%state $\ket{\psi_{00}}$ and vector$\vec{P}_{00}$
vector $\vec{P}_{00}$, whose coordinates $(S_1,S_2)$ are given by the Stokes parameters $S_1 = \mathcal{P}_D - \mathcal{P}_A = 0.848 - 0.152 = 0.696$ and $S_2 = \mathcal{P}_R - \mathcal{P}_L = 0.853 - 0.147 = 0.706$.

Using the same strategy, we can prepare the other states of the simplest scenario, $P_{01},P_{10},$ and $P_{11},$ %\lorenzo{targeting the other states of} $\ket{\psi_{01}}, \ket{\psi_{10}}, \ket{\psi_{11}}$ of Eq.~\eqref{3} 
by setting the angles ($\theta, \alpha$) of $\textrm{QWP}_1$ and $\textrm{HWP}_1$ in the experimental setup shown in Fig.~\ref{Exp_1} to $(\pi/8, -\pi/16), (-\pi/8, \pi/16),$ and $(-\pi/8,5\pi/16)$, respectively. The experimental findings after the tomography block are shown in Fig.~\ref{fig:output_1}. The evaluated Stokes parameters $(S_1, S_2)$ can be calculated using the measured intensities (probabilities) $I_{10}, I_{01},$ and $I_{11}$, corresponding to $(-0.704, 0.696)$, $( 0.716, -0.714)$ and $(-0.694, -0.704)$, respectively. %Note that these experimental findings agree with the vectors' coordinates of the prepared states of Eq.~\eqref{3}.

We can now compute the distances $d(P_{ij}, P_{ij}^{\mathrm{id}})$ defined in Eq.~\eqref{distance} for all prepared polarized states $P_{ij}$.
Based on the experimental data shown in Fig.~\ref{fig:output_1}, we assess whether violations of noncontextuality are witnessed via Pusey's and Marvian's expressions, as well as a violation of parity preservation. The results are summarized in Table~\ref{table1}.

\begin{table}[!ht]
\caption{Summary of results for polarization.}
\label{table1}

\begin{ruledtabular}
\begin{tabular}{cccc}
          & $d({P_{ij},P_{ij}^{\mathrm{id}}})$ & $S(x_{ij},y_{ij})$ & $C^{\mathrm{min}}_{\mathrm{prep}}$ \\
\hline

$P_{00}$ & $0.006$ & $0.734$ & $0.069$     \\ 
$P_{10}$ & $0.007$ & $0.719$ & $0.069$      \\
$P_{01}$ & $0.005$ & $0.750$ & $0.070$      \\ 
$P_{11}$ & $0.007$ & $0.719$ & $0.069$       

\end{tabular}
\end{ruledtabular}

\end{table}

The first column value represents the distance from the prepared (noisy) state to its ideal counterpart, $d({P_{ij},P_{ij}^{\mathrm{id}}})$. Since this distance is equated with the noise parameter $\delta$, the value is used to calculate the second and third column values, which are lower bounds for Pusey's and Marvian's expressions, $S(x_{ij},y_{ij})$ and $C_{\text{prep}}^{\text{min}}$, respectively. These are obtained from Eqs.~\eqref{Pusey3} and \eqref{marvian}. Given the positive values for the expressions $S(x_{ij},y_{ij})$ and $C^{\min}_{\mathrm{prep}}$, the results show that all prepared states witness a violation of Pusey's noncontextuality inequality in Eq.~\eqref{Pusey2} and Marvian's noncontextuality equality,  $C_{\mathrm{prep}}^{\min}=0$. For Pusey's approach, the largest experimental value obtained is for P$_{01}$, namely $S(x_{ij},y_{ij}) \geq 0.75$. This value is close to the theoretical maximum lower bound of $0.82$ attained for the noiseless case in Eq.~\eqref{Pusey3}. Similarly, for Marvian's approach, the largest experimental value obtained is for P$_{01}$ whereby $C^{\min}_{\mathrm{prep}} \geq 0.07$. This value is close to the theoretical maximum lower bound of $0.073$ obtained for the noiseless case in Eq.~\eqref{marvian}. Regarding parity preservation, all prepared states satisfy $d(P_{ij},P^{\text{id}}_{ij}) \leq 0.007$. Consequently, %the conditions $S(x_{ij},y_{ij}) > 0$, $C^{\min}_{\mathrm{prep}} > 0$, and 
$\mathcal{D}^{\min}_{P_+,P_-} > 0$, ensuring that all criteria of classicality are simultaneously violated.

\subsection{Experiment with transverse modes}

\begin{figure}[h!]
    \centering
    \includegraphics[width=0.67\linewidth]{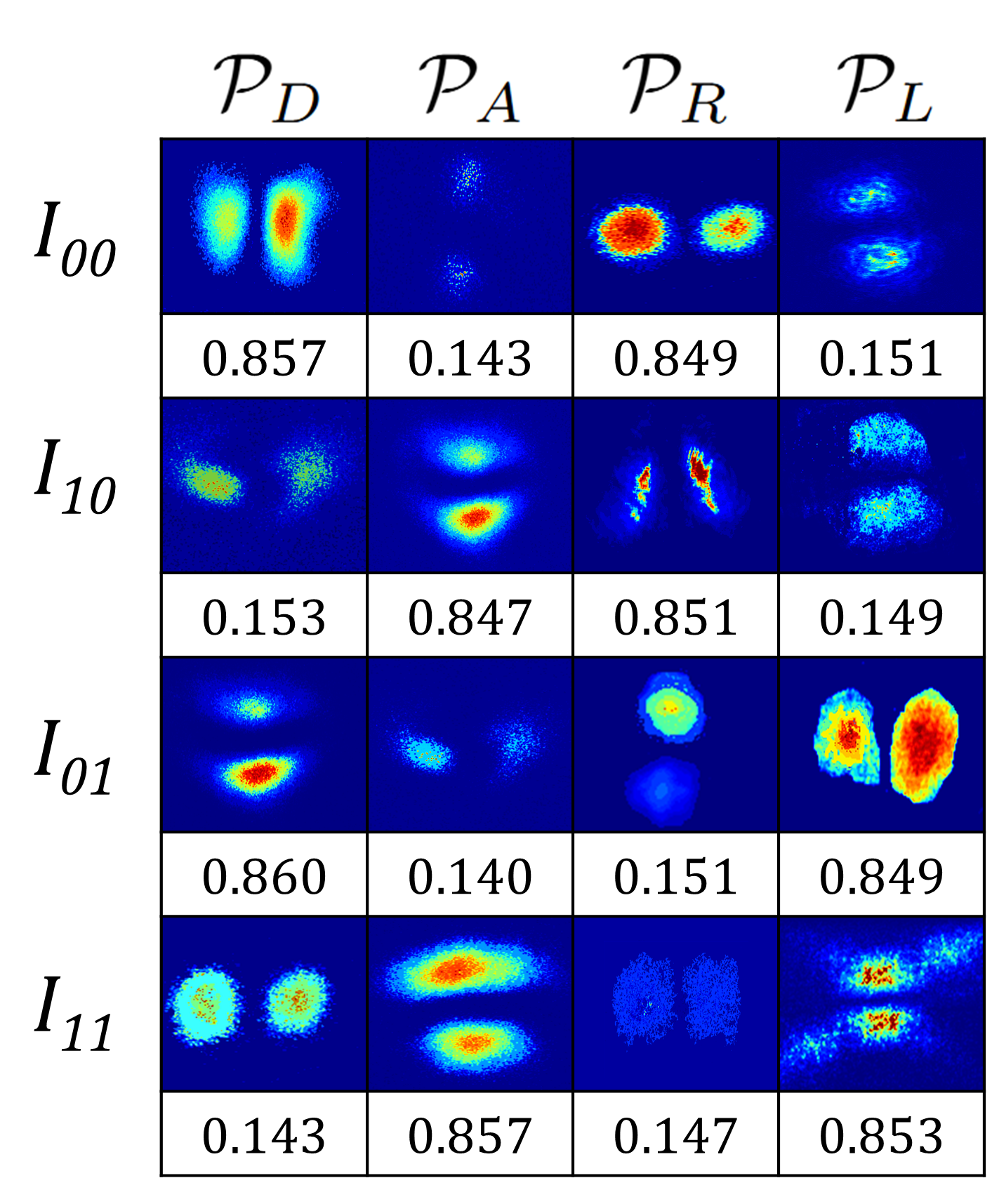}
    \caption{Output images (false color) for prepared first-order Hermite-Gaussian states.} 
    \label{fig:output_2}
\end{figure}

Using the optical setup described in Section~\ref{exp_TM}, we prepare all states $P_{ij}$ of the simplest scenario, aiming to approximate the ideal states $P^{\mathrm{id}}_{ij}$ of Eq.~\eqref{3} as closely as possible.
Fig.~\ref{fig:output_2} shows the measured output intensities for the transverse-mode experiment. 
The horizontal and vertical Hermite-Gaussian modes, $\ket{h}$ and $\ket{v}$, can be clearly distinguished. 
Astigmatic distortions from the cylindrical lenses are also visible. 
The rows in Fig.~\ref{fig:output_2} label the intensities $I_{ij}$ for each prepared state, while the columns label the probabilities associated with the  measurements. The probabilities $\mathcal{P}_D$ and $\mathcal{P}_A$ ($\mathcal{P}_R$ and $\mathcal{P}_L$) are obtained from the intensities measured at output ports, $I_1$ and $I_2$, respectively.
For example, $I_{00}$ is associated with the 
vector $\vec{P}_{00}$, whose coordinates $(S_1,S_2)$ are given by the Stokes parameters 
$S_1 = \mathcal{P}_D - \mathcal{P}_A = 0.857 - 0.143 = 0.714$ and 
$S_2 = \mathcal{P}_R - \mathcal{P}_L = 0.849 - 0.151 = 0.698$.

Using an analogous procedure, we prepare the remaining states of the simplest scenario, $P_{01},P_{10},$ and $P_{11}$, by setting the angles $(\theta, \alpha)$ of $\textrm{CL}_{1,\theta}$ and $\textrm{DP}_{1,\alpha}$ in the experiment in Fig.~\ref{Exp_1} to $(\pi/8, -\pi/16)$, $(-\pi/8, \pi/16)$, and $(-\pi/8, 5\pi/16)$, respectively.  
The measured intensities (probabilities) $I_{01}$, $I_{10}$, and $I_{11}$ yield the Stokes parameters $(S_1, S_2)$ of $(-0.694, 0.702)$, $(0.720, -0.698)$, and $(-0.714, -0.706)$, respectively.
%\st{As happened in the polarization case, these experimental results agree with the coordinates of the vectors of prepared states presented in Eq.}~\eqref{3}. 
As in the previous subsection, we can now compute the distances $d(P_{ij}, P_{ij}^{\mathrm{id}})$ defined in Eq.~\eqref{distance} for all the prepared first-order Hermite-Gaussian states $P_{ij}$. Based on the experimental data shown in Fig.~\eqref{fig:output_2}, we assess whether violations of noncontextuality are witnessed via Pusey's and Marvian's expressions, as well as a violation of parity preservation.  The results are summarized in Table \ref{table2}.

\begin{table}[h]
\caption{Summary of results for transverse modes.}
\label{table2}

\begin{ruledtabular}
\begin{tabular}{cccc}
          & $d({P_{ij},P_{ij}^{\mathrm{id}}})$ & $S(x_{ij},y_{ij})$ & $C^{\mathrm{min}}_{\mathrm{prep}}$ \\
\hline

$P_{00}$ & $0.005$ & $0.750$ & $0.070$        \\
$P_{10}$ & $0.007$ & $0.719$ & $0.069$        \\
$P_{01}$ & $0.007$ & $0.719$ & $0.069$        \\ 
$P_{11}$ & $0.004$ & $0.765$ & $0.071$        

\end{tabular}
\end{ruledtabular}

\end{table}

As before with Table \ref{table1}, the first column value represents the distance between the prepared state and its ideal counterpart: $d({P_{ij},P_{ij}^{\mathrm{id}}})$. This is then used as $\delta$ in calculating the second and third column values, which are lower bounds for Pusey's and Marvian's expressions, $S(x_{ij},y_{ij})$ and $C_{\text{prep}}^{\text{min}}$, respectively. Given the positive values for the expressions $S(x_{ij},y_{ij})$ and $C^{\min}_{\mathrm{prep}}$, the results show that all prepared states witness a violation of Pusey's noncontextuality inequality in Eq.~\eqref{Pusey2} and Marvian's noncontextuality equality, $C_{\mathrm{prep}}^{\min}=0$. For Pusey's approach, the largest experimental value obtained is for P$_{11}$, namely $S(x_{ij},y_{ij}) \geq 0.765$. Similarly, for Marvian's approach, the largest experimental value obtained is for P$_{11}$, whereby $C^{\min}_{\mathrm{prep}} \geq 0.07$1. Regarding parity preservation, all prepared states satisfy $d(P_{ij},P^{\text{id}}_{ij}) \leq 0.007$. Consequently, the prepared Hermite-Gaussian states satisfies the condition $\mathcal{D}^{\min}_{P_+,P_-} > 0$, ensuring that all criteria of classicality are violated.

Note that the output images exhibit astigmatic effects, mainly due to the two pairs of lenses used in the experimental setup. Controlling these lenses and achieving proper mode matching requires careful alignment. Nevertheless, this has not prevented us from performing the projections, which show good agreement with the theoretical predictions even when compared with those obtained for polarization.

\subsection{Experiment under depolarizing noise}

In this subsection, we analyze the results obtained when noise is modeled by a depolarizing channel, as implemented in the experimental setup of Fig.~\ref{Exp_3} and described in Section~\ref{exp_dep_channel}. The depolarizing noise is realized by rotating the half-wave plate $\textrm{HWP}_{1,\beta}$ in steps of one degree, thereby increasing the parameter $a$ in Eq.~\eqref{eq::depolarizing} from zero.

\begin{figure}[htbp]
    \centering

    \begin{subfigure}{0.94\linewidth}
        \centering
        \includegraphics[width=\linewidth]{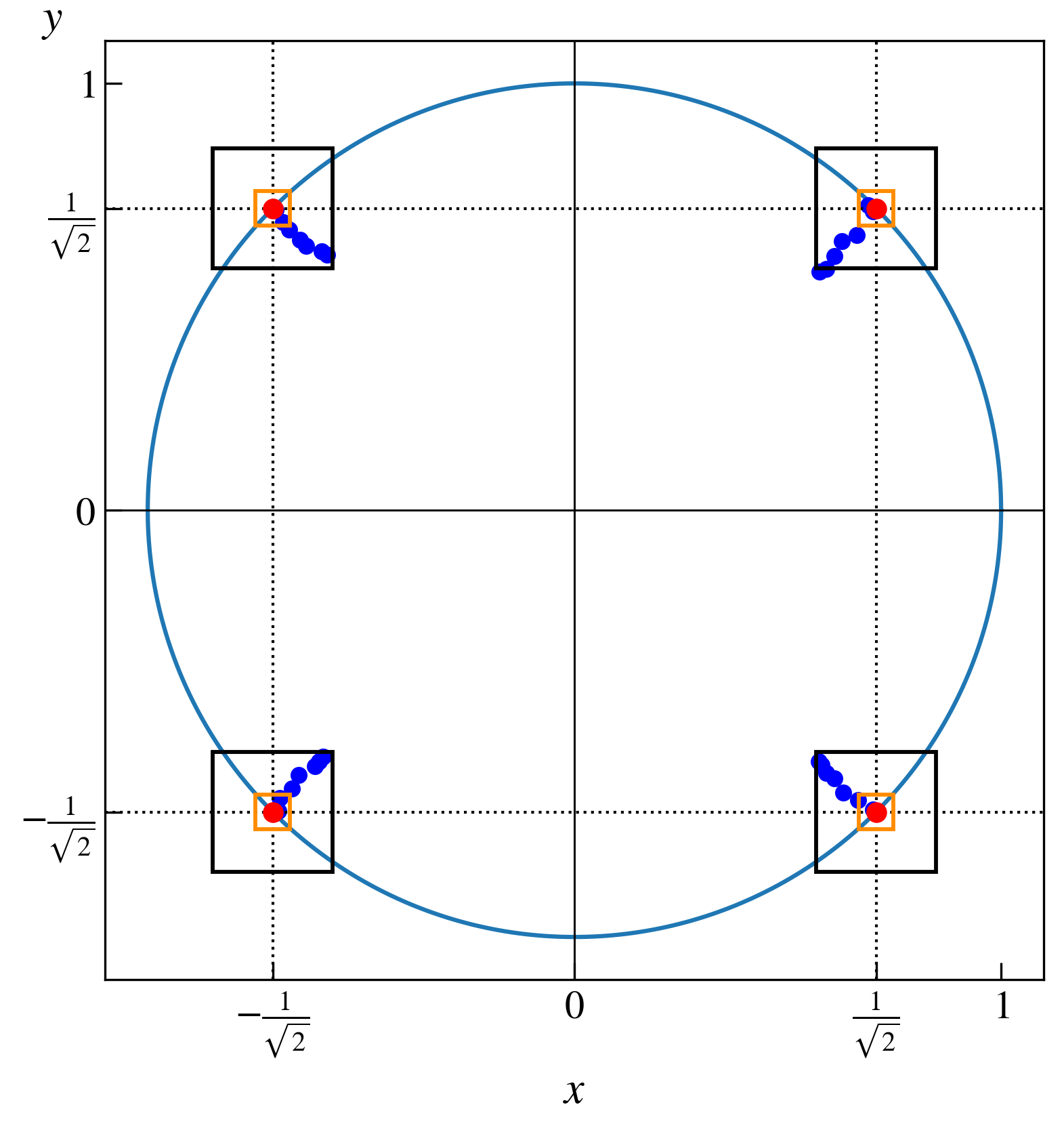}
        \caption{The red dots represent the ideal preparations $P^{\textrm{id}}_{ij}$, while the blue dots represent the experimental points $P_{ij}$. The orange and black squares indicate the thresholds for violation of parity preservation and Pusey’s inequality, respectively.}
        \label{fig::plot_depol}
    \end{subfigure}

    \vspace{0.5cm}

    \begin{subfigure}{0.94\linewidth}
        \centering
        \includegraphics[width=\linewidth]{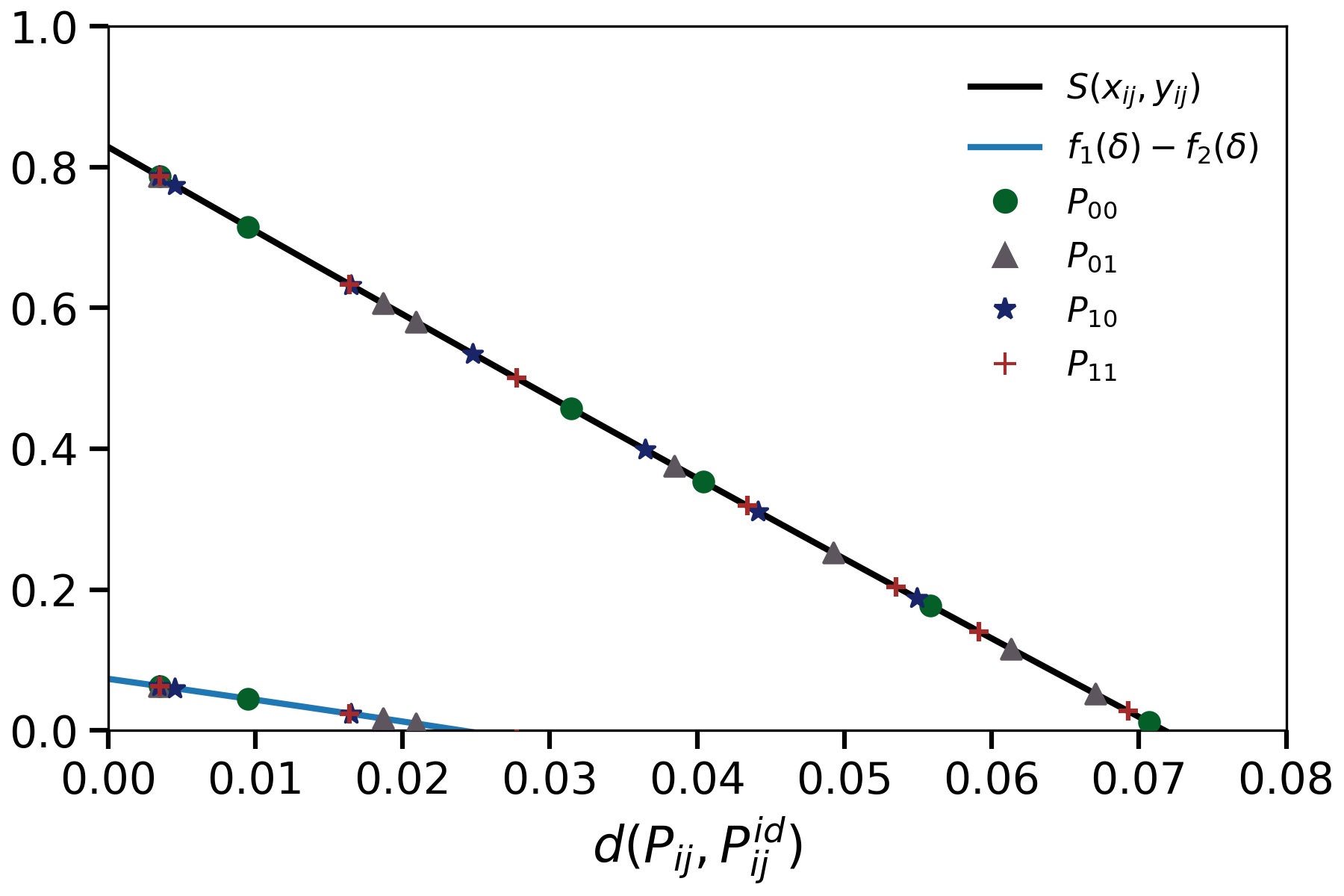}
        \caption{The black solid line shows the bound for Pusey’s expression, $S(x_{ij},y_{ij})$, in Eq.~\eqref{pusey_depol}, and the blue solid line the difference $f_1(\delta) - f_2(\delta)$, which gives the condition for parity preservation described in Eq.~\eqref{marvian_depol} and Fig.~\ref{new_marvian}. The experimental points $P_{ij}$ correspond to the blue dots inside the squares shown in Fig.~\ref{fig::plot_depol}. }
        \label{fig::plot_depol_2}
    \end{subfigure}

    \caption{Experimental results when noise is modeled by a depolarizing channel.}
    \label{fig:simplest_scenario_depolarization}
\end{figure}

Fig.~\ref{fig::plot_depol} shows seven experimental data points (blue dots), together with the corresponding ideal preparation (red dot), for all states $P_{ij}$. The black squares correspond to Pusey’s threshold ($\delta \leq 0.07$), while the orange squares correspond to the threshold for violation of parity preservation ($\delta \leq 0.02$) in the presence of quantum depolarizing noise, as discussed in Section~\ref{section_theo_depol}. We include only seven points, since additional points would exceed the largest threshold associated with Pusey’s approach.

As $a$ increases, the prepared states become more mixed and move toward the center of the Bloch disk, demonstrating that the optical device functions as a depolarizing channel. The distance $d(P_{ij},P_{ij}^{\mathrm{id}})$ between each noisy prepared state and its ideal counterpart is evaluated and used to compute the quantities appearing in Eqs.~\eqref{pusey_depol} and~\eqref{marvian_depol}. For points lying within the orange square, $S(x_{ij}, y_{ij}) > 0$, $C^{\min}_{\mathrm{prep}} > 0$, and $\mathcal{D}^{\min}_{P_{+}, P_{-}} > 0$, implying that all three criteria of classicality are violated. For states inside the black square but outside the orange square, only Pusey’s inequality is violated.

Figure~\ref{fig::plot_depol_2} shows the same experimental points as Fig.~\ref{fig::plot_depol}, now represented with different colors and markers for each $P_{ij}$. The black solid line corresponds to violations of Pusey's inequality, given by Eq.~\eqref{pusey_depol}. The blue solid line corresponds to violations of parity preservation, given by Eq.~\eqref{marvian_depol} and illustrated in Fig.~\ref{new_marvian}. Points within the noise threshold of $d(P_{ij},P_{ij}^{\mathrm{id}}) \leq 0.02$ violate all three criteria of classicality, and correspond to the points within the orange squares in Fig.~\ref{fig::plot_depol}.

\section{Conclusion} \label{sec:conclusions}

In this work, we performed experiments to test the nonclassicality witnesses analyzed in~\cite{Khoshbin2024} for the simplest scenario. We implemented four preparations and a tomographically complete set of two measurements using polarization and first-order Hermite-Gaussian transverse modes of a classical optical beam. Our results are consistent with the theoretical predictions, in the sense that the observed statistics reproduce those expected for the simplest scenario. For Pusey’s approach, we obtain $S(x_{ij}, y_{ij}) \geq 0.719$ for all preparations $P_{ij}$ in both experimental implementations, thereby violating the corresponding noncontextuality inequality. We reach the same conclusion for Marvian’s approach, where we find $C^{\min}_{\mathrm{prep}} \geq 0.069$ for all $P_{ij}$ in both cases. In addition, we observe $\mathcal{D}^{\min}_{P_+,P_-} > 0$ in both implementations, indicating a violation of parity preservation and, consequently, of $\mathrm{BOD}_P$. Overall, for $\delta<0.007$, the observed statistics confirm all the nonclassicality witnesses, in agreement with~\cite{Khoshbin2024}.
We also performed the experiment by modeling noise through a depolarizing channel, testing seven different states for each preparation $P_{ij}$ and confirming the results found in \cite{Khoshbin2024}, where $S(x_{ij}, y_{ij})>0$ for $\delta<0.07$ and where $\mathcal{D}^{\min}_{P_+,P_-} > 0$ for $\delta<0.02$.

Among the novelties of the present work is that we extend experimental tests of generalized contextuality beyond single-photon polarization platforms by using classical light to reproduce the operational statistics of the simplest scenario, using two distinct degrees of freedom: polarization and first-order Hermite--Gaussian transverse modes. Other degrees of freedom, such as path, time, or frequency, are expected to yield analogous results. In addition, we report, to the best of our knowledge, the first experimental investigation of violations of $\mathrm{BOD}_P$.

Our experiment overcomes the issues highlighted in \cite{Mazurek2016} as the problem of noisy measurements -- the idealization that measurements are perfectly noiseless, which is never exactly satisfied in any real experiment -- and the problem of inexact operational equivalences -- the idealization that, in our case, the operational equivalences are exactly as in Eq.~\eqref{noiseless_equ}, which are also never fully realized experimentally. These issues are addressed by the fact that the nonclassicality witnesses introduced in \cite{Khoshbin2024} are robust to noise.
Concerning inexact operational equivalences, each method we consider handles this differently: Pusey's and Marvian's approach take into account the operational equivalences of the experimental preparations, i.e., Eq.~\eqref{op_equ}, while the method based on parity preservation quantifies deviations from the ideal scenario, thereby retaining a direct connection to the noiseless operational equivalence, Eq.~\eqref{noiseless_equ}. 
Regarding noisy measurements, we do not assume that the measurements are noiseless, but only that they form a tomographically complete set. Once this condition is satisfied, then as long as the coordinates of the vectors inferred from the observed output intensities yield $S(x_{ij},y_{ij}) > 0$, $C^{\min}_{\mathrm{prep}} > 0$, and $\mathcal{D}^{\min}_{P_+,P_-} > 0$, the experimental statistics are guaranteed to be inconsistent with any ontological model satisfying preparation noncontextuality and $BOD_P$.

One may then wonder whether it is reasonable to assume that the measurements we perform in the diagonal and circular bases are tomographically complete. This assumption is crucial, as every assessment of operational equivalence between two preparations -- from which one deduces the consequences of noncontextuality -- relies on comparing their statistics for a tomographically complete set of measurements. In our experiment, the diagonal and circular measurements are independently calibrated and sufficiently distinct, which we take as a practically justified assumption of tomographic completeness, while residual imperfections are treated as bounded noise.
Of course, as has been argued previously in \cite{Mazurek2016,Mazurek2021,Grabowecky2022,Schmid2025}, one can never be absolutely certain that the implemented measurements form a tomographically complete set, since it is impossible to test all possible measurements. Nevertheless, these works have proposed methods to increase confidence that the measurements are indeed tomographically complete.
In \cite{Mazurek2021}, the authors introduced a general probabilistic theory (GPT) tomography scheme that simultaneously characterizes preparations and measurements from a large set of experimental data, without prior assumptions, later termed bootstrap GPT tomography and applied to a three-dimensional photonic system in \cite{Grabowecky2022}. More recently, \cite{Schmid2025} showed that the relevant notion is relative tomographic completeness, whereby assessments of noncontextuality require only that the implemented preparations and measurements be tomographically complete with respect to each other.
We leave for future research the application of these techniques to the experimental scenario considered in this work.

We recall that the states and measurements of the simplest scenario shown in Fig.~\ref{fig:noiseless_case} underpin computational advantages in tasks such as $2 \rightarrow 1$ quantum random access codes~\cite{Ambainis1999,Spekkens2008} and the CHSH* game~\cite{Henaut2018,Faleiro2024}, which constitute some of the simplest communication primitives enabling semi-device-independent certification of nonclassicality. Although our implementation uses classical light, it reproduces the operational statistics of this scenario, which are precisely the statistics required for these protocols. An interesting future direction is to experimentally investigate scenarios involving eight preparations and three tomographically complete measurements, with applications to the $3 \rightarrow 1$ parity-oblivious multiplexing task~\cite{Spekkens2008}, which is likewise powered by preparation contextuality and is relevant for quantum communication protocols.  A further natural step is to consider scenarios in which noise is modeled by channels other than the depolarizing channel, such as the dephasing channel, as studied in Refs.~\cite{Rossi2023,Fonseca2025}.  Finally, we emphasize that optical implementations based on intense laser beams, as in the present experiment, can also be realized in the single-photon regime or with attenuated laser beams by changing the light source and detection scheme, and are particularly well suited to integrated and fiber-based platforms.

\begin{acknowledgments}

We would like to thank the financial support from the Brazilian funding agencies Conselho Nacional de Desenvolvimento Cientifíco e Tecnológico (CNPq), Fundação Carlos Chagas Filho de Amparo à Pesquisa do Estado do Rio de Janeiro (FAPERJ), Coordenação de Aperfeiçoamento de Pessoal de Nível Superior (CAPES), National Institute for Science and Technology for Quantum Devices (INCT-QD / CNPq, Grant No. 408783/2024-9), and for Applied Quantum Computing (INCT-CQA/CNPq process No. 408884/2024-0). LC acknowledges funding from the Horizon Europe project FoQaCiA, GA no.101070558.
\end{acknowledgments}

\bibliography{apssamp}% Produces the bibliography via BibTeX.

\end{document}